\documentclass[pra,aps,twocolumn,epsfig,superscriptaddress,showpacs]{revtex4}
\usepackage{mathrsfs}
\usepackage{amsfonts}
\usepackage{amsmath,graphicx,epsfig,color,CJK}
\usepackage{pifont,bbding,amssymb}
\usepackage[hypertex]{hyperref}

\begin{document}
\title{Multipartite Leggett-type Inequalities}
\author{Dong-Ling Deng}
\affiliation{Department of Physics and MCTP, University of Michigan, Ann Arbor, Michigan 48109, USA}
\affiliation{Center for Quantum Information, IIIS, Tsinghua University, Beijing, China}
\affiliation{Centre for Quantum
Technologies, National University of Singapore, 3 Science Drive 2,
Singapore 117543}

\author{Chunfeng Wu}
\affiliation{Centre for Quantum
Technologies, National University of Singapore, 3 Science Drive 2,
Singapore 117543}

\author{Jing-Ling Chen}
\email{chenjl@nankai.edu.cn}
\affiliation{Centre for Quantum
Technologies, National University of Singapore, 3 Science Drive 2,
Singapore 117543}
\affiliation{Theoretical Physics Division, Chern Institute of
Mathematics, Nankai University, Tianjin 300071, People's Republic of
China}

\author{C. H. Oh}
\email{phyohch@nus.edu.sg}
\affiliation{Centre for Quantum
Technologies, National University of Singapore, 3 Science Drive 2,
Singapore 117543}
\affiliation{Department of Physics, National University of Singapore, 2 Science Drive 3,
Singapore 117542}

\date{\today}

\begin{abstract}
We use two different approaches to derive multipartite Leggett-type inequalities, which are generalizations of the two-qubit Leggett-type inequality obtained in [Nature
Phys. \textbf{4}, 681 (2008)]. The first approach is based on the assumption that the probability distributions should be non-negative. The second approach is based on a very simple algebraic equation and is, to some extent, easier than the first approach. Although these inequalities might not
be the optimal ones in the sense that their quantum violations may not be the strongest, our results make the first step of generalizing
Leggett-type inequality to multi-qubit systems and provide the
possibility to experimentally test non-local realism in such
systems. Moreover, the two approaches here may shed new light on the challenging problem of obtaining stronger multipartite Leggett-type inequalities.
\end{abstract}


\maketitle

The concept of physical realism suggests that the results of
observations are determined by the intrinsic properties of a
physical system and independent of measurements~\cite{1935Einstein}.
This concept has taken root in classical physics and its
significance goes far beyond science. Quantum mechanics (QM),
however, challenges this concept in a very deep way. In $1964$, Bell published his celebrated inequality based on Einstein-Podolsky-Rosen's notion of local realism (LR) \cite{Bell1964}. It was shown that any local realistic theory should obey Bell's inequality, while it
can be violated easily in QM. Today this brilliant inequality has
had various generations~\cite{BI-Continous,2qudits-BI,Bell-Inequalities,Multipartite-BI,2004Chen} and many experiments have been carried out to test these Bell's
inequalities~\cite{BI-Experiments}. Notwithstanding some loophole
problems~\cite{2007Cabello}, these experiments overwhelmingly
support QM and show violations of Bell inequalities, and thus rendering
local hidden variable models untenable.

Then, should non-local realism hold water? To answer this
fundamental question, Leggett in $2003$ made a significant step by
proposing an alternative non-local hidden variable (NLHV) model that
was proved to be at variance with quantum
predictions~\cite{2003Leggett}. He introduced a new inequality based
on this model and showed that this inequality can be violated by
quantum correlations. Recently, a series of experiments have been
carried out to test such
model~\cite{2007Leggett-Experiments,2008Branciard,2007Paterek,2007Branciard}. These
experiments again favor QM, casting doubt on the validity of
non-local realism. Nevertheless, all these works only concern with
two-qubit systems. Up to now, we still lack multipartite
Leggett-type inequalities that can be used to experimentally test
nonlocal realism in multi-qubit systems.


In this paper, we generalize the Leggett-type inequality to
multi-qubit systems by using two different approaches. The first approach follows a recent work by Branciard and his collaborators~\cite{2008Branciard}. While the second approach is based on a very simple equation and is a little bit easier. These
inequalities are all satisfied by Leggett's NLHV model and violated by QM. Thus, our
results have paved the way for experimental test of nonlocal realism
in multi-qubit systems. However, a drawback of these inequalities is that the quantum violations do not scale with the system size as the violations of Bell inequalities do~\cite{Multipartite-BI,2004Laskowski}. In fact, the maximal quantum violations are the same as for two-qubit case.

For convenience and simplicity, we first focus at three-qubit
system and the generalization to $n$-qubit case is obvious. Consider
a common experimental scenario: three observers, denoted by A, B,
and C (or Alice, Bob and Charlie), perform experiments with settings labeled by
$\mathbf{a}$, $\mathbf{b}$ and $\mathbf{c}$, respectively. Their
outcomes are denoted by $\alpha$, $\beta$ and $\gamma$
$(\alpha,\beta,\gamma=\pm1)$. For the qubit case, $\mathbf{a}$,
$\mathbf{b}$ and $\mathbf{c}$ are vectors on the Poincar\'{e} sphere
and are independently and freely chosen by Alice, Bob and Charlie. Originally, Leggett's NLHV model is based on pairs of photons and three main assumptions~\cite{2007Leggett-Experiments,2003Leggett}:
(i) Realism. All measurement outcomes are predetermined and
independent of the measurements. (ii) Definite polarization. Physical states are statistical mixtures of subensembles. Within each subensemble, every photon in the pair has definite polarization. (iii) Malus's law. Within each subensemble, local
marginals should obey Malus's law. Based on these assumptions, Leggett was able to derive an inequality that was shown to be violated in QM. In our simplified formulism, assumption (i)
means that the outcomes of each observable is predetermined by some set of hidden variables $\lambda$, polarization parameters $\mathbf{u}$, $\mathbf{v}$ and $\mathbf{s}$ ($\mathbf{u}$, $\mathbf{v}$ and $\mathbf{s}$ are all unit vectors), and some set of other possible non-local parameters $\chi$ (here for simplicity, we choose these non-local parameters to be measurement settings in space-like separated regions~\cite{2007Leggett-Experiments}). Mathematically, we have $\alpha=\alpha(\lambda,\mathbf{u}, \mathbf{v},\mathbf{s},\mathbf{a},\mathbf{b},\mathbf{c})$, $\beta=\beta(\lambda,\mathbf{u}, \mathbf{v},\mathbf{s},\mathbf{a},\mathbf{b},\mathbf{c})$ and $\gamma=\gamma(\lambda,\mathbf{u}, \mathbf{v},\mathbf{s},\mathbf{a},\mathbf{b},\mathbf{c})$. Note that here the locality requirement in local hidden variable theory is explicitly removed. This implies a big difference in deriving Bell inequalities and Leggett-type inequalities. For assumption (ii) and (iii), we denote the polarization distribution function of subensembles and the probability distribution of $\lambda$ in each subensemble by $\texttt{D}(\mathbf{u}, \mathbf{v},\mathbf{s})$ and $\rho_{\mathbf{u}, \mathbf{v},\mathbf{s}}(\lambda)$, respectively. Then the expectation values of measurements for each subensemble are given by average of the measurement outcomes over the probability distribution $\rho_{\mathbf{u}, \mathbf{v},\mathbf{s}}(\lambda)$. For example, we have $\overline{\alpha\beta}(\mathbf{u},\mathbf{v},\mathbf{s})=\int d\lambda\rho_{\mathbf{u}, \mathbf{v},\mathbf{s}}(\lambda)\alpha(\lambda,\mathbf{a},\mathbf{b},\mathbf{c})\beta(\lambda,\mathbf{a},\mathbf{b},\mathbf{c})$ and $\overline{\alpha\beta\gamma}(\mathbf{u},\mathbf{v},\mathbf{s})=\int d\lambda\rho_{\mathbf{u}, \mathbf{v},\mathbf{s}}(\lambda)\alpha(\lambda,\mathbf{a},\mathbf{b},\mathbf{c})\beta(\lambda,\mathbf{a},\mathbf{b},\mathbf{c})
\gamma(\lambda,\mathbf{a},\mathbf{b},\mathbf{c})$. Furthermore, the assumption of Malus's law indicates $\overline{\alpha}(\mathbf{u})=\int d\lambda \rho_{\mathbf{u}, \mathbf{v},\mathbf{s}}(\lambda) \alpha(\lambda,\mathbf{u}, \mathbf{v},\mathbf{s},\mathbf{a},\mathbf{b},\mathbf{c})=\mathbf{u}\cdot\mathbf{a}$, and similarly $\overline{\beta}(\mathbf{v})=\mathbf{v}\cdot\mathbf{b}$,  $\overline{\gamma}(\mathbf{s})=\mathbf{s}\cdot\mathbf{c}$.  The assumption (ii) also implies that the physically measurable correlation functions are given by averaging the expectation values over the subensemble distribution $\texttt{D}(\mathbf{u}, \mathbf{v},\mathbf{s})$. For instance, the three body correlation function is given by:
\begin{eqnarray}
\mathcal{Q}=\langle\alpha\beta\gamma\rangle=\int d\mathbf{u}d\mathbf{v}d\mathbf{s}\texttt{D}(\mathbf{u}, \mathbf{v},\mathbf{s})\overline{\alpha\beta\gamma}(\mathbf{u},\mathbf{v},\mathbf{s}).
\end{eqnarray}

After the introduction of the three major assumptions and the basic notions, now let us derive the three-qubit Leggett-type inequality using two approaches. The first approach follows the recent result of Ref.~\cite{2008Branciard}. To this end, note that the conditional probability distribution
$\texttt{P}(\alpha,\beta,\gamma|\mathbf{a},\mathbf{b},\mathbf{c})$
can be expressed as:
\begin{eqnarray}
\texttt{P}(\alpha,\beta,\gamma|\mathbf{a},\mathbf{b},\mathbf{c})=
\int d\kappa\texttt{D}(\kappa)\texttt{P}_{\kappa}
(\alpha,\beta,\gamma|\mathbf{a},\mathbf{b},\mathbf{c}),
\end{eqnarray}
where we use a single parameter $\kappa$ to denote $(\mathbf{u}, \mathbf{v},\mathbf{s})$ and  $\texttt{P}_{\kappa}
(\alpha,\beta,\gamma|\mathbf{a},\mathbf{b},\mathbf{c})$ is the conditional probability of subensemble labeled by $\kappa$. In
the qubit case, $\texttt{P}_{\kappa}
(\alpha,\beta,\gamma|\mathbf{a},\mathbf{b},\mathbf{c})$ can be decomposed as:
\begin{eqnarray}\label{Peta}
\texttt{P}_{\kappa}(\alpha,\beta,\gamma|\mathbf{a},\mathbf{b},\mathbf{c})
&=&\frac{1}{8}[1+\alpha\mathcal
{L}^{\texttt{A}}_{\kappa}(\mathbf{a},\mathbf{b},\mathbf{c})+\beta\mathcal
{L}^{\texttt{B}}_{\kappa}(\mathbf{a},\mathbf{b},\mathbf{c})\nonumber\\
&+&\gamma\mathcal
{L}^{\texttt{C}}_{\kappa}(\mathbf{a},\mathbf{b},\mathbf{c})+\alpha\beta\mathcal
{L}^{\texttt{AB}}_{\kappa}(\mathbf{a},\mathbf{b},\mathbf{c})\nonumber\\
&+&\alpha\gamma\mathcal
{L}^{\texttt{AC}}_{\kappa}(\mathbf{a},\mathbf{b},\mathbf{c})+\beta\gamma\mathcal
{L}^{\texttt{BC}}_{\kappa}(\mathbf{a},\mathbf{b},\mathbf{c})\nonumber\\
&+&\alpha\beta\gamma\mathcal
{L}^{\texttt{ABC}}_{\kappa}(\mathbf{a},\mathbf{b},\mathbf{c})]
\end{eqnarray}
One advantage of this expression is that it enables one to clearly
distinguish the marginals and the correlation coefficient as
discussed in Ref.~\cite{2008Branciard}.  From the above Eq.~(\ref{Peta}), it is obvious that
$\mathcal{L}^{\texttt{A}}_{\kappa}(\mathbf{a},\mathbf{b},\mathbf{c})$,
$\mathcal{L}^{\texttt{B}}_{\kappa}(\mathbf{a},\mathbf{b},\mathbf{c})$,
$\mathcal{L}^{\texttt{C}}_{\kappa}(\mathbf{a},\mathbf{b},\mathbf{c})$,
$\mathcal{L}^{\texttt{AB}}_{\kappa}(\mathbf{a},\mathbf{b},\mathbf{c})$,
$\mathcal{L}^{\texttt{AC}}_{\kappa}(\mathbf{a},\mathbf{b},\mathbf{c})$,
$\mathcal{L}^{\texttt{BC}}_{\kappa}(\mathbf{a},\mathbf{b},\mathbf{c})$, and
$\mathcal{L}^{\texttt{ABC}}_{\kappa}(\mathbf{a},\mathbf{b},\mathbf{c})$ have
their physical meaning of average within subensemble $\kappa$. For instance, Alice, Bob and
Charlie's marginals can be respectively expressed as:
\begin{eqnarray}\label{LA}
\mathcal{L}^{\texttt{A}}_{\kappa}(\mathbf{a},\mathbf{b},\mathbf{c})&=&\sum_{\alpha,\beta,\gamma}
\alpha\texttt{P}_{\kappa}(\alpha,\beta,\gamma|\mathbf{a},\mathbf{b},\mathbf{c})=\overline{\alpha}(\mathbf{u}),\nonumber\\
\mathcal{L}^{\texttt{B}}_{\kappa}(\mathbf{a},\mathbf{b},\mathbf{c})&=&\sum_{\alpha,\beta,\gamma}
\beta\texttt{P}_{\kappa}(\alpha,\beta,\gamma|\mathbf{a},\mathbf{b},\mathbf{c})=\overline{\beta}(\mathbf{v}),\nonumber\\
\mathcal{L}^{\texttt{C}}_{\kappa}(\mathbf{a},\mathbf{b},\mathbf{c})&=&\sum_{\alpha,\beta,\gamma}
\gamma\texttt{P}_{\kappa}(\alpha,\beta,\gamma|\mathbf{a},\mathbf{b},\mathbf{c})=\overline{\gamma}(\mathbf{s}).\nonumber
\end{eqnarray}
Similarly, the two-qubit correlation coefficients read:
\begin{eqnarray}\label{LAB}
\mathcal{L}^{\texttt{AB}}_{\kappa}(\mathbf{a},\mathbf{b},\mathbf{c})&=&\sum_{\alpha,\beta,\gamma}
\alpha\beta\texttt{P}_{\kappa}(\alpha,\beta,\gamma|\mathbf{a},\mathbf{b},\mathbf{c})=\overline{\alpha\beta},\nonumber\\
\mathcal{L}^{\texttt{AC}}_{\kappa}(\mathbf{a},\mathbf{b},\mathbf{c})&=&\sum_{\alpha,\beta,\gamma}
\alpha\gamma\texttt{P}_{\kappa}(\alpha,\beta,\gamma|\mathbf{a},\mathbf{b},\mathbf{c})=\overline{\alpha\gamma},\nonumber\\
\mathcal{L}^{\texttt{BC}}_{\kappa}(\mathbf{a},\mathbf{b},\mathbf{c})&=&\sum_{\alpha,\beta,\gamma}
\beta\gamma\texttt{P}_{\kappa}(\alpha,\beta,\gamma|\mathbf{a},\mathbf{b},\mathbf{c})=\overline{\beta\gamma},\nonumber
\end{eqnarray}
and the three-qubit correlation coefficient reads:
\begin{eqnarray}\label{LABC}
\mathcal{L}^{\texttt{ABC}}_{\kappa}(\mathbf{a},\mathbf{b},\mathbf{c})=\sum_{\alpha,\beta,\gamma}
\alpha\beta\gamma\texttt{P}_{\kappa}(\alpha,\beta,\gamma|\mathbf{a},\mathbf{b},\mathbf{c})=\overline{\alpha\beta\gamma}.\nonumber
\end{eqnarray}
As in Ref.~\cite{2008Branciard}, we also only concentrate on
correlations satisfying the so called no-signaling condition, which
implies the independence of marginals on other observer's inputs. For the marginals on Alice, Bob or Charlie's side, the condition is already certified by Malus's law~\cite{2007Leggett-Experiments}. The no-signaling condition also indicates three more equations: $\mathcal{L}^{\texttt{AB}}_{\kappa}(\mathbf{a},\mathbf{b},\mathbf{c})=\mathcal{L}^{\texttt{AB}}_{\kappa}(\mathbf{a},\mathbf{b})$, $\mathcal{L}^{\texttt{AC}}_{\kappa}(\mathbf{a},\mathbf{b},\mathbf{c})=\mathcal{L}^{\texttt{AC}}_{\kappa}(\mathbf{a},\mathbf{c})$, and $\mathcal{L}^{\texttt{BC}}_{\kappa}(\mathbf{a},\mathbf{b},\mathbf{c})=\mathcal{L}^{\texttt{BC}}_{\kappa}(\mathbf{b},\mathbf{c})$.
The probability distributions
$\texttt{P}_{\kappa}(\alpha,\beta,\gamma|\mathbf{a},\mathbf{b},\mathbf{c})$
should be non-negative for all $\alpha,\beta,\gamma=\pm1$. This leads to eight inequalities:
\begin{subequations}
\begin{eqnarray}
1+\mathcal {L}^{\texttt{A}}_{\kappa}+\mathcal
{L}^{\texttt{B}}_{\kappa}+\mathcal
{L}^{\texttt{C}}_{\kappa}+\mathcal {L}^{\texttt{AB}}_{\kappa}
+\mathcal {L}^{\texttt{AC}}_{\kappa}+\mathcal
{L}^{\texttt{BC}}_{\kappa}+\mathcal
{L}^{\texttt{ABC}}_{\kappa}\geq0,\label{RPPP}\\
1+\mathcal {L}^{\texttt{A}}_{\kappa}+\mathcal
{L}^{\texttt{B}}_{\kappa}-\mathcal
{L}^{\texttt{C}}_{\kappa}+\mathcal {L}^{\texttt{AB}}_{\kappa}
-\mathcal {L}^{\texttt{AC}}_{\kappa}-\mathcal
{L}^{\texttt{BC}}_{\kappa}-\mathcal
{L}^{\texttt{ABC}}_{\kappa}\geq0,\label{RPPM}\\
1+\mathcal {L}^{\texttt{A}}_{\kappa}-\mathcal
{L}^{\texttt{B}}_{\kappa}+\mathcal
{L}^{\texttt{C}}_{\kappa}-\mathcal {L}^{\texttt{AB}}_{\kappa}
+\mathcal {L}^{\texttt{AC}}_{\kappa}-\mathcal
{L}^{\texttt{BC}}_{\kappa}-\mathcal
{L}^{\texttt{ABC}}_{\kappa}\geq0,\label{RPMP}\\
1+\mathcal {L}^{\texttt{A}}_{\kappa}-\mathcal
{L}^{\texttt{B}}_{\kappa}-\mathcal
{L}^{\texttt{C}}_{\kappa}-\mathcal {L}^{\texttt{AB}}_{\kappa}
-\mathcal {L}^{\texttt{AC}}_{\kappa}+\mathcal
{L}^{\texttt{BC}}_{\kappa}+\mathcal
{L}^{\texttt{ABC}}_{\kappa}\geq0,\label{RPMM}\\
1-\mathcal {L}^{\texttt{A}}_{\kappa}+\mathcal
{L}^{\texttt{B}}_{\kappa}+\mathcal
{L}^{\texttt{C}}_{\kappa}-\mathcal {L}^{\texttt{AB}}_{\kappa}
-\mathcal {L}^{\texttt{AC}}_{\kappa}+\mathcal
{L}^{\texttt{BC}}_{\kappa}-\mathcal
{L}^{\texttt{ABC}}_{\kappa}\geq0,\label{RMPP}\\
1-\mathcal {L}^{\texttt{A}}_{\kappa}+\mathcal
{L}^{\texttt{B}}_{\kappa}-\mathcal
{L}^{\texttt{C}}_{\kappa}-\mathcal {L}^{\texttt{AB}}_{\kappa}
+\mathcal {L}^{\texttt{AC}}_{\kappa}-\mathcal
{L}^{\texttt{BC}}_{\kappa}+\mathcal
{L}^{\texttt{ABC}}_{\kappa}\geq0,\label{RMPM}\\
1-\mathcal {L}^{\texttt{A}}_{\kappa}-\mathcal
{L}^{\texttt{B}}_{\kappa}+\mathcal
{L}^{\texttt{C}}_{\kappa}+\mathcal {L}^{\texttt{AB}}_{\kappa}
-\mathcal {L}^{\texttt{AC}}_{\kappa}-\mathcal
{L}^{\texttt{BC}}_{\kappa}+\mathcal
{L}^{\texttt{ABC}}_{\kappa}\geq0,\label{RMMP}\\
1-\mathcal {L}^{\texttt{A}}_{\kappa}-\mathcal
{L}^{\texttt{B}}_{\kappa}-\mathcal
{L}^{\texttt{C}}_{\kappa}+\mathcal {L}^{\texttt{AB}}_{\kappa}
+\mathcal {L}^{\texttt{AC}}_{\kappa}+\mathcal
{L}^{\texttt{BC}}_{\kappa}-\mathcal
{L}^{\texttt{ABC}}_{\kappa}\geq0,\label{RMMM}
\end{eqnarray}
\end{subequations}
From the inequalities (\ref{RPPP}-\ref{RMMM}), it is easy to obtain:
\begin{eqnarray}\label{LabcSingleM}
|\mathcal{L}^{\texttt{A}}_{\kappa}(\mathbf{a})\pm\mathcal{L}^{\texttt{BC}}_{\kappa}(\mathbf{b},\mathbf{c})|\leq1\pm
\mathcal{L}^{\texttt{ABC}}_{\kappa}(\mathbf{a},\mathbf{b},\mathbf{c}).
\end{eqnarray}
Now let us consider the case that Alice has two measurement settings $\mathbf{a}$ and $\mathbf{a}'$, while Bob and Charlie have only one measurement setting $\mathbf{b}$ and $\mathbf{c}$, respectively. Then we can obtain four inequalities from inequality~(\ref{LabcSingleM}), two for $(\mathbf{a},\mathbf{b},\mathbf{c})$ and two for $(\mathbf{a}',\mathbf{b},\mathbf{c})$. Combining these four inequalities and using the triangle inequality, one arrives at:
\begin{eqnarray}\label{LABCU}
|\mathcal{L}^{\texttt{ABC}}_{\kappa}(\mathbf{a},\mathbf{b},\mathbf{c})\pm
\mathcal{L}^{\texttt{ABC}}_{\kappa}(\mathbf{a}',\mathbf{b},\mathbf{c})|+|\mathbf{u}\cdot\mathbf{a}\mp \mathbf{u}\cdot\mathbf{a}'|\leq 2.
\end{eqnarray}
Here the equations $\mathcal{L}^{\texttt{A}}_{\kappa}(\mathbf{a})=\mathbf{u}\cdot\mathbf{a}$ and $\mathcal{L}^{\texttt{A}}_{\kappa}(\mathbf{a}')=\mathbf{u}\cdot\mathbf{a}'$ indicated by Malus's law are used. To derive three-qubit Leggett's inequality, let's consider three $4$-tuple settings $(\mathbf{a}_i,\mathbf{a}'_i,\mathbf{b}_i,\mathbf{c}_i) (i=1,2,3)$. The angle between all pairs $(\mathbf{a}_i,\mathbf{a}'_i)$ is the same $\theta$ and $(\mathbf{a}'_i-\mathbf{a}_i)=2\sin\frac{\theta}{2}\mathbf{e}_i$, where $\{\mathbf{e}_1,\mathbf{e}_2,\mathbf{e}_3\}$ form an orthogonal basis. Then combining the three corresponding inequalities~(\ref{LABCU}) and the fact that $\sum_{i=1}^3|\mathbf{e}_i\cdot \mathbf{u}|\geq1$, one obtains $\sum_{i=1}^3|\mathcal{L}^{\texttt{ABC}}_{\kappa}(\mathbf{a}_i,\mathbf{b}_i,\mathbf{c}_i)+
\mathcal{L}^{\texttt{ABC}}_{\kappa}(\mathbf{a}'_i,\mathbf{b}_i,\mathbf{c}_i)|+2|\sin\frac{\theta}{2}|\leq6$. Doing an integration over $\texttt{D}(\mathbf{u}, \mathbf{v},\mathbf{s})$ and using the fact that $\int  d\mathbf{u}d\mathbf{v}d\mathbf{s}\texttt{D}(\mathbf{u}, \mathbf{v},\mathbf{s})=1$, we obtain a three-qubit Leggett-type inequality:
\begin{eqnarray}\label{ThreeQubit-LTI}
\mathcal {I}_3=\sum_{i=1}^3|\mathcal {Q}_{iii}+\mathcal {Q}_{i'ii}|
+2|\sin\frac{\theta}{2}|\leq6,
\end{eqnarray}
where $\mathcal {Q}_{jii}=\int d\kappa\texttt{D}(\kappa)\mathcal{L}^{\texttt{ABC}}_{\kappa}(\mathbf{a}_j,\mathbf{b}_i,\mathbf{c}_i) (\mathbf{a}_j=\mathbf{a}_i \texttt{ if } j=i; {\rm and} \;\mathbf{a}_j=\mathbf{a}'_i \texttt{ if } j=i')$ is the physically measurable correlation function.  Inequality~(\ref{ThreeQubit-LTI}) is our final three-qubit Leggett-type inequality. This inequality is a generalization of the
two-quibt inequality obtained in Ref~.\cite{2008Branciard}. To obtain this inequality, we used the eight inequalities~(\ref{RPPP}-\ref{RMMM}). In fact, there is an alternative approach, which is a little bit easier to derive three-qubit Leggett-type inequality, if we note the fact that the following equation is valid for any $\alpha,\beta,\gamma=\pm1$:
\begin{eqnarray}\label{SimpleEq}
|\alpha\pm \beta\gamma|\mp \alpha\beta\gamma=1
\end{eqnarray}
Doing an integration $\int d\lambda\rho_{\mathbf{u}, \mathbf{v},\mathbf{s}}(\lambda)$ for both side of Eq.~(\ref{SimpleEq}), one can easily get inequality~(\ref{LabcSingleM}) and hence inequality~(\ref{ThreeQubit-LTI}) by the same method. It is worthwhile to underline that the second approach does not require the eight inequalities coming from the positivity of probability distribution. It only relies on a very simple algebraic Eq.~(\ref{SimpleEq}). Thus, it is, to some extent, a little bit easier.

\begin{figure}
\includegraphics[width=85mm]{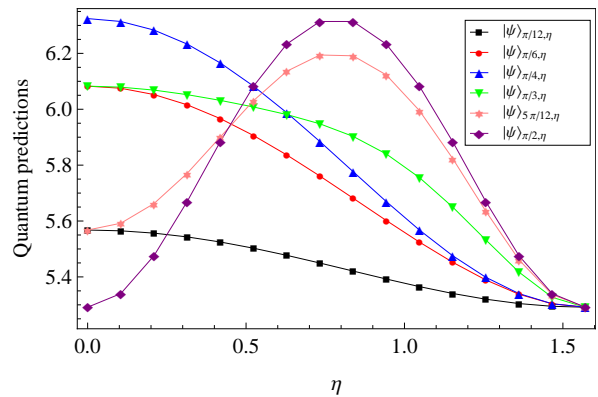}\\
\caption{(Color online) Numerical results of quantum violations of inequality (\ref{ThreeQubit-LTI}) by the generalized W states $|\psi\rangle^{\rm W}_3=\sin\xi\cos\eta|100\rangle+\sin\xi\sin\eta|010\rangle+\cos\xi|001\rangle$.
Quantum predictions are plotted versus the variation of $\eta$ with $\xi=\pi/12,\xi=\pi/6,\xi=\pi/4,\xi=\pi/3,\xi=5\pi/12,\xi=\pi/2$, and $\eta$ is from $0$ to $\pi/2$.
Inequality (\ref{ThreeQubit-LTI}) is violated for the states whose quantum predictions are larger than $6$.
 }\label{quantum-violations}
\end{figure}

Quantum prediction for the correlation functions is $\mathcal{Q}_{ijk}=\langle \mathbf{a}_i\cdot\vec{\sigma}\otimes\mathbf{b}_j\cdot\vec{\sigma}\otimes\mathbf{c}_k\cdot\vec{\sigma}\rangle$,
where $\vec{\sigma}$ are Pauli matrices. For three qubits, we consider two types of entangled states: GHZ state and W state.
We first look at the GHZ state $|\psi\rangle^{\rm GHZ}_3=1/\sqrt{2}(|000\rangle+|111\rangle)$.
By choosing the measurement settings as
$\mathbf{a}_1=(\cos\frac{\theta}{2},-\sin\frac{\theta}{2},0)$,
$\mathbf{a}_2=(0,\cos\frac{\theta}{2},-\sin\frac{\theta}{2})$,
$\mathbf{a}_3=(-\sin\frac{\theta}{2},\cos\frac{\theta}{2},0)$,
$\mathbf{a}'_1=(\cos\frac{\theta}{2},\sin\frac{\theta}{2},0)$,
$\mathbf{a}'_2=(0,\cos\frac{\theta}{2},\sin\frac{\theta}{2})$,
$\mathbf{a}'_3=(\sin\frac{\theta}{2},\cos\frac{\theta}{2},0)$,
$\mathbf{b}_1=\mathbf{c}_1=(1,0,0)$, and
$\mathbf{b}_2=\mathbf{b}_3=\mathbf{c}_2=\mathbf{c}_3=(\frac{\sqrt{2}}{2},\frac{\sqrt{2}}{2},0)$, it is found that the quantum value of
the left-hand side of inequality (\ref{ThreeQubit-LTI}) is
$\mathcal{I}^Q_4=6(\cos\frac{\theta}{2}+\frac{1}{3}\sin\frac{\theta}{2})$, which is larger than $6$ when $0<\theta<4\arctan\frac{1}{3}$. Moreover, noting that $6(\cos\frac{\theta}{2}+\frac{1}{3}\sin\frac{\theta}{2})=2\sqrt{10}\sin(\frac{\theta}{2}+\theta_0)$, the maximal quantum violation is $2\sqrt{10}$ and it occurs when $\theta=2\arctan\frac{1}{3}\approx 36.9^o$. The maximal violation of inequality
(\ref{ThreeQubit-LTI}) by the GHZ state is the same as that of the inequality given in Ref. \cite{2008Branciard} by singlet
state. We then consider the quantum violation of inequality (\ref{ThreeQubit-LTI}) by generalized W states
$|\psi\rangle^{\rm W}_3=\sin\xi\cos\eta|100\rangle+\sin\xi\sin\eta|010\rangle+\cos\xi|001\rangle$.
In Fig. \ref{quantum-violations}, we show the numerical results of quantum violation by the family of generalized W states for the
cases $\xi=\pi/12,\xi=\pi/6,\xi=\pi/4,\xi=\pi/3,\xi=5\pi/12,\xi=\pi/2$. It is clear that for some values of $\xi$ and $\eta$, the inequality (\ref{ThreeQubit-LTI})
is violated by $|\psi\rangle^{\rm W}_3$. Furthermore, as pointed out in Ref.\cite{3qubitstate}, an arbitrary three-qubit pure state can be written in terms of five
parameters as (up to local unitary transformations):
\begin{eqnarray}\label{arbistate}
|\psi\rangle_3&=&\sqrt{\mu_0}|000\rangle+\sqrt{\mu_1}e^{i\phi}|100\rangle+\sqrt{\mu_2}|101\rangle\nonumber \\
&&+\sqrt{\mu_3}|110\rangle+\sqrt{\mu_4}|111\rangle,
\end{eqnarray}
with $\mu_i\geq 0$, $\sum_i\mu_i=1$, and $0\leq \phi\leq \pi$. For $|\psi\rangle_3$, our numerical results show that the inequality~(\ref{ThreeQubit-LTI}) is also violated for some region of the parameters and the maximal violation happens at $\mu_0=\mu_4=\frac{1}{2}$ and $\mu_1=\mu_2=\mu_2=0$, corresponding to the GHZ state.

Our three-qubit Leggett-type inequality~(\ref{ThreeQubit-LTI}) can be easily generalized to the general $n$-qubit case. For all these inequalities, Alice has six measurement settings while other parties each has three settings. In general, the $n$-qubit Leggett-type inequality reads:
\begin{eqnarray}\label{nQubit-LI}
\mathcal {I}_n=\sum_{i=1}^3|\mathcal {Q}_{ii\cdots i}+\mathcal {Q}_{i'i\cdots i}|
+2|\sin\frac{\theta}{2}|\leq6,
\end{eqnarray}
where $\mathcal {Q}_{ji\cdots i}$ are physically measurable $n$-qubit correlation functions. The inequality~(\ref{nQubit-LI}) is violated for various quantum entangled states. Its maximal violation happens for $n$-qubit GHZ state $|\psi\rangle_n^{\rm GHZ}=\frac{1}{\sqrt{2}}(|0\cdots0\rangle+|1\cdots1\rangle)$ and the maximal violation is $2\sqrt{10}$, too. In fact, for four-qubit case, we numerically calculated the quantum violations for all the pure entangled states and find that the maximal quantum violation for four-qubit Leggett-type inequality~(\ref{nQubit-LI}) is indeed $2\sqrt{10}$.

To summarize, we have derived multipartite Leggett-type inequalities by using two different approaches. These inequalities are generalizations
of the two-qubit Leggett-type inequality obtained in Ref.~\cite{2008Branciard}. The maximal violation of these inequalities is always $2\sqrt{10}$, and thus it does not scale with the system size. This shows a shortcome of our generalized inequalities. Of course, to acquire larger quantum violation and hence smaller experimental visibility threshold, one can think about increasing the number of measurement settings, as discussed in Ref.~\cite{2008Branciard}. However, increasing the number of settings will make the experimental tests more difficult. To obtain stronger multipartite Leggett-type inequalities with few settings is a very challenging problem. We hope that the two approaches presented in this paper will stimulate other works in this direction. Moreover, so far all the experimental tests of Leggett-type inequalities suffer from the ``detection inefficient" loophole. A promising loophole free experimental scheme is to test these Leggett-type inequalities in topologically ordered systems that host non-Abelian anyons. Candidate systems including quantum Hall liquid~\cite{2004Xia}, rotating Bose condensates~\cite{2001Cooper}, quantum spin systems~\cite{2004Freedman} as well as Aharonov-Casher phase~\cite{ours}. The entangled states can be obtained by braiding non-Abelian anyons~\cite{2010Deng}, while the measurements is implemented by fusing anyons together. We will investigate this scheme subsequently.

J.L.C. is supported by National Basic Research Program (973 Program)
of China under Grant No. 2012CB921900 and NSF of China (Grant Nos.
10975075 and 11175089). This work is also partly supported by the IARPA
MUSIQC program, the ARO and the AFOSR MURI program,
and partly by National Research Foundation and Ministry of Education, Singapore
(Grant No. WBS: R-710-000-008-271).

\end{document}